# Strength of the dominant scatterer in graphene on silicon oxide

Jyoti Katoch[1,2], Duy Le[1], Simranjeet Singh[1], Rahul Rao, Talat S. Rahman[1] and Masa Ishigami[1,2*]

[1]Department of Physics, University of Central Florida, Orlando, FL 32816-2385, USA

[2]Nanoscience Technology Center, University of Central Florida, Orlando, FL 32816-2385, USA

*To whom correspondence should be addressed. E-mail: ishigami@ucf.edu.

## Abstract

A large variability of carrier mobility of graphene-based field effect transistors hampers graphene science and technology. We determine the scattering strength of the dominant scatterer responsible for the variability of graphene-based transistors on silicon oxide. The strength of the scatterer is found to be more consistent with charged impurities than with resonant impurities.

Graphene possesses exceptionally high room temperature carrier mobility, which can be utilized for high frequency field effect transistors (FETs) [1], sensors [2] and transparent conductors [3, 4], as well as for exploration of two-dimensional physics. Yet, its mobility is known to vary by more than an order of magnitude without any control at the synthesis or fabrication level and even gentle treatments such as annealing are known to modify the mobility of graphene without affecting its atomic structure [5], severely affecting the utility of graphene in science and technology.

The origin of the observed variability has been attributed to charged [6] and resonant impurities [7-11]. Each type of scatterer possesses a different scattering potential and, therefore, strength. Charged impurities near graphene are observed to affect its mobility as $1/\mu_{FE} = (2\times10^{-16})\ n_{charged}$ V sec [12], and vacancies, model resonant impurities, will affect mobility as $1/\mu_{FE} = (1\times10^{-15})\ n_{resonant}$ V sec [13]. However thus far, the number density of the dominant scatterer has not been correlated to mobility. The typical density of charged impurities on $SiO_2$ of $50\times10^{10}$/cm$^2$ is consistent with the maximum observed mobility on $SiO_2$, but such match does not explain the variability. In addition, a large variability is still observed for the mobility values of graphene on hexagonal boron nitride [14] and suspended graphene [15] although these device configurations should minimize the impact from the substrate-bound charged impurities. Finally, gate-dependent resistivity of graphene can be fitted to theoretical expectations [6, 9, 10, 16, 17] by choosing correct impurity density, potential, and locations, but such analysis remains speculative because these adjustable parameters cannot be independently confirmed.

In this paper, we correlate the number of scatterers to field effect mobility of graphene devices on $SiO_2$, which is the most commonly used substrate for graphene science and technology, and find that the scattering strength of the dominant scatterer responsible for the observed variability is more consistent with that of charged impurities.

Graphene field effect transistors (FETs) are fabricated from mechanically exfoliated graphene on $SiO_2$ using conventional e-beam lithography [18]. Devices are cleaned down to atomic scale by annealing in $Ar/H_2$ at 350 °C for 3 hours [19] and annealed subsequently in ultra-high vacuum at above 400 K for approximately 12 hours prior to each experiment. Devices are exposed to atomic hydrogen generated using a hydrogen cracker [20] at temperatures ranging from 8 to 15 K. Hydrogen is introduced by means of a leak valve resulting in a constant hydrogen flux of approximately $3.7 \times 10^{12}$ H /cm²/sec at devices. After graphene is saturated with hydrogen, its desorption from graphene is measured as a function of temperature. We perform desorption measurements on three devices with a range of mobility in order to understand the differences in desorption characteristics. We also perform measure-anneal cycles on two devices to understand the impact of annealing on the variability. After each measurement, devices are dehydrogenated at above 400 K for longer than 12 hours to recover hydrogen-free graphene.

Figure 1(a) shows an example of the impact of atomic hydrogen on the transport properties of graphene with the initial mobility of 14,000 cm²/V sec for electrons and 9,700 cm²/V sec for holes. Previously, we explored the impact of atomic hydrogen on graphene on $SiO_2$ in detail [21], where we observed that the number of the dominant scatter is proportional to the number of atomic hydrogen on graphene at the saturation coverage. Upon dosing with atomic hydrogen, the gate voltage at which the conductivity is minimized ($V_{min}$) shifts to more negative values. The amount of voltage shift to $V_{min}$ induced by hydrogen, $V_{shift}$, is proportional to the resistivity added by atomic hydrogen adsorbed on the surface. This indicates that $V_{shift}$ is proportional to number of atomic hydrogen and adsorbed hydrogen atoms behave as isolated scattering sites. The effect of atomic hydrogen is observed to saturate and $V_{saturation}$, $V_{shift}$ at the saturation limit, was found to be inversely proportional to the initial carrier mobility indicating that the dominant scatterer has enhanced affinity towards hydrogen. Finally, atomic hydrogen is weakly bound to graphene and is readily desorbed as shown in Figure 1(b). $V_{shift}$ decreases and eventually approaches zero as devices are annealed at 400 K. In this paper, we determine charge transfer from atomic hydrogen to the scattering sites to correlate the mobility to the number of the dominant scatterer, in order to find the scattering strength of the scatterer and identify the cause for the variability.

The interaction strength between hydrogen and graphene can be measured by observing desorption characteristics. Figure 2a shows the temperature dependence of $V_{shift}$ as the representative graphene device shown in Figure 1 is warmed from 15 to 300 K. Hydrogen desorbs from graphene nearly continuously in this temperature range, indicating a large range of desorption energies. Figure 2b shows desorption characteristics of hydrogen from another graphene device with the initial mobility of 7,100 $cm^2$/V sec for electrons and

4,900 cm$^2$/V sec for holes. The desorption characteristics observed from the two devices are apparently similar, but there are some differences.

Correlating the variability in the desorption characteristics to the variability in the initial mobility reveals that only hydrogen with a particular range of desorption energies "counts" the dominant scatterer. Figure 3(a) shows that $V_{shift}$ induced by warming from 100 to 200 K is inversely proportional to the initial mobility while $V_{shift}$ induced by warming outside this particular temperature range is not correlated to the initial mobility as shown in Figures 3(b-c). As such, we conclude that only hydrogen atoms, desorbing between 100 and 200 K, count the dominant scatterer. The same scatterer is also responsible for the variations in device performance observed after thermal annealing as results from two annealing cycles also fall on the same slope.

A linear fit to $V_{shift}$ induced by warming from 100 to 200 K shows that $V_{shift(100-200K)}$= $(27000\pm10000)\times1/\mu_{FE}$ + $(-0.58\pm1.22)$ volts for electrons and $(22000\pm5800)\times1/\mu_{FE}$ + $(-1.16\pm1.00)$ volts for electrons, where $\mu_{FE}$ is measured in cm$^2$/V sec. Fitting similarly for $1/\mu_{FE}$, we obtain: $1/\mu_{FE}$= $[(2.58\pm0.97) \times 10^{-5} V_{shift(100-200K)} + (4.9\pm0.3) \times 10^{-5}]$ V sec/cm$^2$ for electrons and $1/\mu_{FE}$= $[(3.4\pm1.0) \times 10^{-5} V_{shift(100-200K)} + (7.4\pm2.7) \times 10^{-5}]$ V sec/cm$^2$. The value for mobility with $V_{shift(100-200K)}$ at zero is 20,000±11,000 cm$^2$/V sec for electrons and 13,500±4,900 cm$^2$/V sec for holes, corresponding well to the maximum observed mobility of graphene on silicon oxide. Interestingly, these zero intercept values also indicate that other scattering mechanisms, different from the source for the observed variability, are responsible for limiting the maximum mobility to 10,000~30,000 cm$^2$/V sec on SiO$_2$.

The determined temperature range for desorption of the special hydrogen reveals the interaction strength between hydrogen and the scattering sites. For 1$^{st}$ order desorption, $E_{desorption} = kT_{max}[\ln(fT_{max}/\beta) - 3.64]$ where $k$ is the Boltzmann constant, $T_{max}$ is the temperature of maximum desorption rate in K, $f$ is the attempt frequency in Hz, and $\beta$ is the heating rate in K/sec [22]. While the attempt frequency is difficult to determine directly, it is often assumed that $f = kT/h$, where $T$ is the surface temperature and $h$ is the Planck's constant. Using $T_{max}$ ranging from 100 to 200 K, we find the desorption energy ranges approximately from 300 to 600 meV for hydrogen counting scattering sites. These desorption energies are close to the value of 600 meV measured for desorption of chemisorbed hydrogen on graphite [23].

Finding charge donated per chemisorbed hydrogen is necessary to determine the relationship between the mobility and the number of scattering sites and, therefore, the scattering strength of the dominant scatterer. Charge donated by chemisorbed hydrogen, which desorb below room temperature has not been determined experimentally. In order to calculate charge donated by these special hydrogen atoms, total energy calculations are carried out within the density functional theory (DFT), using the super cell method with plane-wave basis set and the projector-augmented wave (PAW)[24, 25] technique as implemented in the Vienna Ab-initio Simulation Package (VASP) [26, 27]. We account for the van der Waals interaction by using DFT-D3 method [28], which we consider to be appropriate here based on an earlier comparative study [29] of the different methods for calculating binding energy of organic molecules on graphene. Our simulation supercell consists of a (7×7) graphene layer, a hydrogen atom on one side, and a vacuum of 20 Å to separate normal periodical image. We sample the Brillouin Zone with one point at the zone-center, which is found to be sufficient for supercells containing a large number of atoms as here. We set the cutoff energy for plane-wave expansion at 500 eV. The geometry of chemisorbed H on graphene is found by placing the H atom close to the surface (~1.2 Å from a C atom) and allowing all atoms in the super cell to relax until forces acting on each atom is smaller than 0.01 eV/Å. The charge transfer from chemisorbed hydrogen to graphene is determined to be 0.06 electrons using the Bader analysis [30].

The calculated charge transfer and the data presented in Figure 3(a) yield $1/\mu_{FE} = (2.0 \pm 0.8) \times 10^{-17} n + (4.9 \pm 2.7) \times 10^{-5}$ V sec / cm$^2$ for electrons and $1/\mu_{FE} = (2.7 \pm 0.8) \times 10^{-17} n + (7.4 \pm 2.7) \times 10^{-5}$ V sec / cm$^2$ for electrons, where n is the number of adsorption sites which is the number of the dominant scatterer since adsorbed hydrogen is known not to cluster. The observed scattering strength is 50 times smaller than expected for the resonant scatterers and it is 10 times smaller than expected for charged impurities. Therefore, our data favor charged impurities as the dominant scatterer responsible for the variability in mobility of graphene on SiO$_2$.

In conclusion, we used atomic hydrogen to correlate the number of scatterers to field effect mobility of graphene devices on $SiO_2$. The relationship between the number of scatterers and the initial field effect mobility suggests that charged impurities are responsible for the variability observed in the performance of graphene-based field effect transistors on $SiO_2$.

**Acknowledgements**: This work is supported by National Science foundation grant no. 0955625. DFT work by DL and TSR is support in part by National Science Foundation grant CHE-1310327 and is carried out at the Stokes Advanced Research Computing Center (Stokes ARCC) at the University of Central Florida. Authors thank Eduardo Mucciolo for helpful discussions and Enrique del Barco and Nina Orlovskaya for sharing instruments in their lab.

# Figure 1

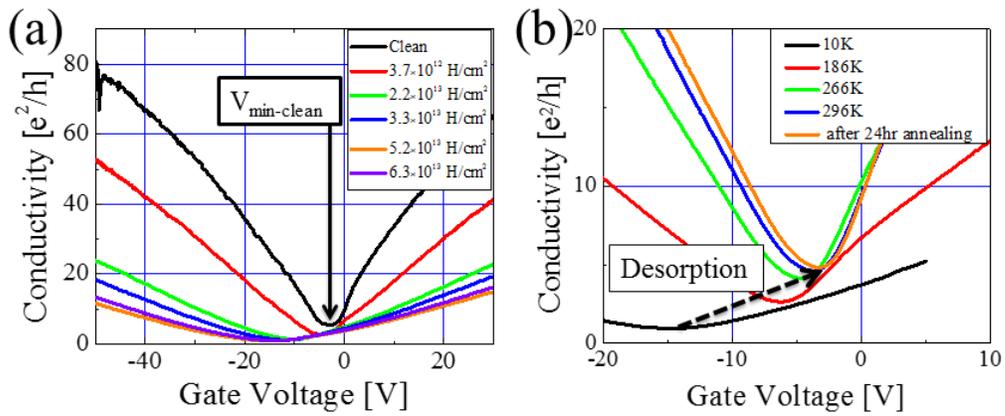

Figure 1(a) Conductivity as a function of gate voltage for a representative device. Black curve shows the transport in undoped clean graphene sample. (b) The temperature dependent measurements of hydrogenated graphene. As the temperature is increased from 10K to 400K the dehydrogenation of graphene is evident from the decrease in $V_{shift}$.

Figure 2

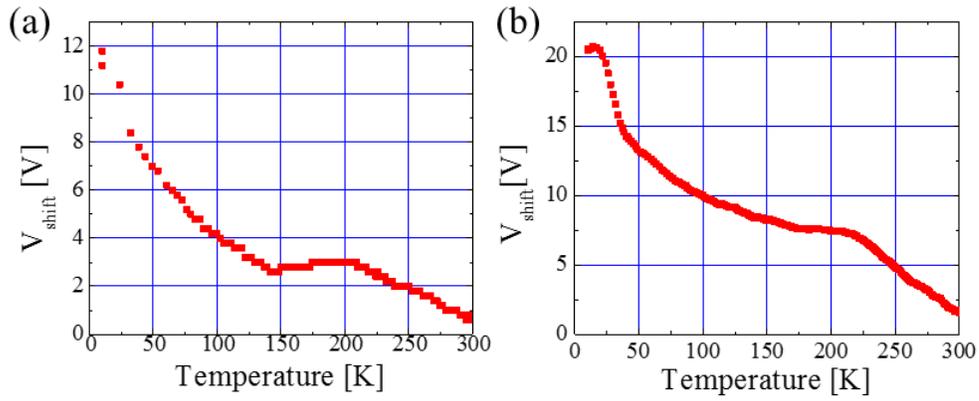

Figure 2 (a), (b) $V_{shift}$ as a function of temperature for two different graphene devices. The heating rate of 4.2 to 1.0 K/min was used for both these samples.

Figure 3

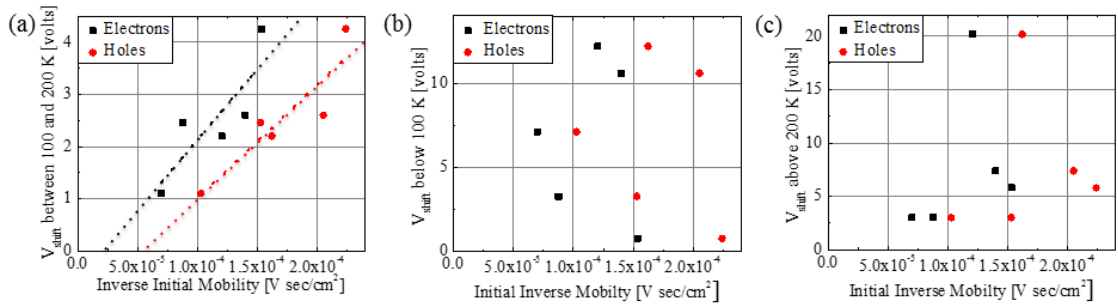

Figure 3 $V_{shift}$ in specific temperature ranges as a function of inverse initial mobility of all the measured graphene samples showing (a) $V_{shift}$ in temperature range 100-200K, (b) $V_{shift}$ below 100 K and (c) $V_{shift}$ above 200 K.